\documentclass[twoside]{LCWS11}
\usepackage[latin1]{inputenc}
\usepackage[dvips]{graphicx,epsfig,color}
\usepackage{wrapfig,rotating}
\usepackage{amssymb,amsmath,array}
\usepackage{cite}
\include{paperdef}
\graphicspath{{figs/}}

\begin{document}

\thispagestyle{empty}
\setcounter{page}{0}
\def\thefootnote{\fnsymbol{footnote}}

\begin{flushright}
\mbox{}
\end{flushright}

\vspace{1cm}

\begin{center}

{\large\sc {\bf Implications of SUSY Searches at the LHC for the ILC}}
\footnote{Invited talk given by S.H.\ at the {\em LCWS 2011}, 
September 2011, Granada, Spain}

\vspace{1cm}

{\sc 
S.~Heinemeyer
\footnote{
email: Sven.Heinemeyer@cern.ch}%
}

\vspace*{1cm}

{\it
Instituto de F\'isica de Cantabria (CSIC-UC), 
Santander,  Spain 

}
\end{center}

\vspace*{0.2cm}

\BC {\bf Abstract} \EC
Global frequentist fits to the CMSSM and NUHM1 using the
{\tt MasterCode} framework are updated to include the public results of
searches for 
supersymmetric signals using $\sim 1$/fb of LHC data
recorded by ATLAS and CMS and $\sim 0.3$/fb of data recorded by LHCb.
We also include the constraints imposed by the 
electroweak precision and B-physics observables, 
the cosmological dark matter density
and the XENON100 search for spin-independent dark matter scattering.
Finally we also investigate the impact of a possible Higgs signal around
$125 \gev$.
The various constraints set new bounds on the parameter space of the
CMSSM and the NUHM1. We discuss the impact of the new results from SUSY
and Higgs searches for these bounds and analyze the impact for a future
liner $e^+e^-$ collider.

\def\thefootnote{\arabic{footnote}}
\setcounter{footnote}{0}

\newpage


\title{Implications of SUSY Searches at the LHC for the ILC}
\author{S.~Heinemeyer
\vspace{.3cm}\\
Instituto de F\'isica de Cantabria (CSIC)
E-39005 Santander, Spain
}

\maketitle

\begin{abstract}
Global frequentist fits to the CMSSM and NUHM1 using the
{\tt MasterCode} framework are updated to include the public results of
searches for 
supersymmetric signals using $\sim 1$/fb of LHC data
recorded by ATLAS and CMS and $\sim 0.3$/fb of data recorded by LHCb.
We also include the constraints imposed by the 
electroweak precision and B-physics observables, 
the cosmological dark matter density
and the XENON100 search for spin-independent dark matter scattering.
Finally we also investigate the impact of a possible Higgs signal around
$125 \gev$.
The various constraints set new bounds on the parameter space of the
CMSSM and the NUHM1. We discuss the impact of the new results from SUSY
and Higgs searches for these bounds and analyze the impact for a future
liner $e^+e^-$ collider.

\end{abstract}

\section{Introduction}

The {\tt MasterCode}
collaboration~\cite{mc1,mc2,mc3,mc35,mc4,mc5,mc6,mc7,mc75,mc-web} 
has reported the results of global fits to
pre-LHC data~\cite{mc1,mc2,mc3,mc35,mc4} (see \citere{pre-LHC} for other
works) as well as including LHC 2010 data~\cite{mc5,mc6,mc7,mc75,mc-web} (see
\citere{post-LHC} for other works) in the frameworks of simplified
variants of the minimal supersymmetric 
extension of the Standard Model (MSSM)~\cite{HK}
with universal supersymmetry-breaking mass parameters at the GUT scale.
We consider a class of models in which R-parity is conserved and the
lightest supersymmetric particle (LSP), assumed to be the lightest
neutralino $\neu{1}$~\cite{EHNOS},  
provides the cosmological cold dark matter~\cite{Komatsu:2010fb}.
The specific models studied have included the constrained MSSM
(CMSSM) with parameters $m_0$, $m_{1/2}$ and $A_0$ denoting common
scalar, fermionic and trilinear soft supersymmetry-breaking parameters
at the GUT scale, and $\tb$ denoting the ratio of the 
two vacuum expectation values of the two Higgs fields. Other models
studied include a model in which common supersymmetry-breaking
contributions to the Higgs masses are allowed to be 
non-universal (the NUHM1), a very constrained model in which trilinear
and bilinear soft supersymmetry-breaking parameters are related (the
VCMSSM), and minimal supergravity (mSUGRA) in which the gravitino
mass is required to be the same as the universal soft
supersymmetry-breaking scalar mass before renormalization (see
\citere{mc7} for an extensive list of references for those models)

The impressive increase in the accumulated LHC luminosity combined with
the rapid analyses of LHC data by the ATLAS~\cite{ATLASsusy,ATLASHA},
CMS~\cite{CMSsusy,CMSHA,CMSbmm} and LHCb Collaborations~\cite{LHCbbmm} 
have been included in the analysis presented in \citere{mc7}. 
Most recently, the ATLAS and CMS Collaborations have presented
preliminary updates of their results for the search for a SM-like Higgs
boson with $\sim 5$/fb of data~\cite{Dec13}. These results may be
compatible with a SM-like Higgs boson around $\Mh \simeq 125 \gev$,
and they have been included in the analysis presented in \citere{mc75}.

Here we review the results of \citeres{mc7,mc75}, which are focused on the
CMSSM and the NUHM1. We discuss their
impact on the physcis prospects for a future linear $e^+e^-$ collider, 
in particular the ILC, with a center-of-mass energy around $1 \tev$ (the
ILC(1000)).


\section{The {\tt MasterCode} framework}
\label{sec:mc}

We define a global $\chi^2$ likelihood function, which combines all
theoretical predictions with experimental constraints (except the latest
LHC SUSY and Higgs searches):
\begin{align}
\chi^2 (\equiv \chi^2_{\rm org}) &=
  \sum^N_i \frac{(C_i - P_i)^2}{\sigma(C_i)^2 + \sigma(P_i)^2}
+ \sum^M_i \frac{(f^{\rm obs}_{{\rm SM}_i}
              - f^{\rm fit}_{{\rm SM}_i})^2}{\sigma(f_{{\rm SM}_i})^2}
\nonumber \\[.2em]
&+ {\chi^2(\Mh) + \chi^2(\br(B_s \to \mu\mu))}~.
\label{eqn:chi2}
\end{align} 
Here $N$ is the number of observables studied, $C_i$ represents an
experimentally measured value (constraint) and each $P_i$ defines a
prediction for the corresponding constraint that depends on the
supersymmetric parameters.
The experimental uncertainty, $\sigma(C_i)$, of each measurement is
taken to be both statistically and systematically independent of the
corresponding theoretical uncertainty, $\sigma(P_i)$, in its
prediction. We denote by
$\chi^2(\Mh)$ and $\chi^2(\bmm)$ the $\chi^2$
contributions from the two measurements for which only one-sided
bounds were previously included.

We stress that the three SM parameters
$f_{\rm SM} = \{\Delta\alpha_{\rm had}, \mt, \MZ \}$ are included as fit
parameters and allowed to vary with their current experimental
resolutions $\sigma(f_{\rm SM})$. We do not
include $\alpha_s$ as a fit parameter, 
which would have only a minor impact on the analysis.

Formulating the fit in this fashion has the advantage that the
$\chi^2$ probability, $P(\chi^2, N_{\rm dof})$,
properly accounts for the number of degrees of freedom, $N_{\rm dof}$,
in the fit and thus represents a quantitative and meaningful measure for
the ``goodness-of-fit.'' In previous studies \cite{mc1},
$P(\chi^2, N_{\rm dof})$ has been verified to have a flat distribution,
thus yielding a reliable estimate of the confidence level for any particular
point in parameter space. 
All confidence levels for selected model parameters are
performed  by scanning over the desired parameters while  
minimizing the $\chi^2$ function with respect to all other model parameters. 
The function values where $\chi^2(x)$ is found to be equal to 
$\chi^2_{min}+ \Delta \chi^2$ determine the confidence level
contour. For two-dimensional parameter scans we use 
$\Delta \chi^2 =2.23 (5.99)$ to determine the 68\%(95\%) confidence
level contours. 
Only experimental constraints are imposed when deriving confidence level
contours, without any arbitrary or direct constraints placed on model
parameters themselves.
This leads to robust and statistically meaningful
estimates of the total 68\% and 95\% confidence levels,
which may be composed of multiple separated contours.

The experimental constraints used in our analyses are listed in
Table~2 in \cite{mc7}. 
It should be noted that we use of the $e^+ e^-$ determination of the SM
contribution to \gmt~\cite{newDavier}, 
$a_\mu^{\rm SUSY} = (30.2 \pm 8.8) \times 10^{-10}$.
Indeed, the
\gmt\ hint has even strengthened with the convergence of the previously
discrepant SM calculations using low-energy $e^+ e^-$ and $\tau$ decay
data~\cite{newDavier,Jegerlehner}. 

The numerical evaluation of the frequentist likelihood function
using the constraints has been performed with the 
{\tt MasterCode}~\cite{mc1,mc2,mc3,mc35,mc4,mc5,mc6,mc7,mc75,mc-web},
which includes the following theoretical codes. For the RGE running of
the soft SUSY-breaking parameters, it uses
{\tt SoftSUSY}~\cite{Allanach:2001kg}, which is combined consistently
with the codes used for the various low-energy observables. 
At the electroweak scale we have included various codes:
{\tt FeynHiggs}~\cite{FeynHiggs} is used for the evaluation of the Higgs
masses and $a_\mu^{\rm SUSY}$ (see also
\cite{Moroi:1995yh,Degrassi:1998es,Heinemeyer:2003dq,Heinemeyer:2004yq}).
For flavor-related observables we use 
{\tt SuFla}~\cite{SuFla} as well as 
{\tt SuperIso}~\cite{SuperIso}, and
for the electroweak precision data we have included 
a code based on~\cite{Svenetal}.
Finally, for dark-matter-related observables, 
{\tt MicrOMEGAs}~\cite{MicroMegas} and
{\tt SSARD}~\cite{SSARD} 
have been used.
We made extensive use of the SUSY Les Houches
Accord~\cite{SLHA} 
in the combination of the various codes within the {\tt MasterCode}.

The {\tt MasterCode} framework is such that new observables can easily
be incorporated via new `afterburners', as we discuss below for the
LHC$_{\rm 1/fb}$ constraints. 
We use a Markov Chain Monte Carlo (MCMC) approach to sample the parameter
spaces of supersymmetric models. The sampling is based on the
Metropolis-Hastings algorithm, with a multi-dimensional Gaussian
distribution as proposal density. The width of this distribution is
adjusted during the sampling, so as to keep the 
MCMC acceptance rate between 20\% and 40\% in order to ensure efficient
sampling. 
It should be noted that we do not make use of the sampling density to infer 
the underlying probability distribution.  
The results of \citeres{mc7,mc75} are based on a basic
resampling of the CMSSM with $7\,10^7$ points and a resampling of the NUHM1
with $7\,10^7$ additional points, both extending up to 
$m_0, m_{1/2} = 4 \tev$. 
We check that the afterburners we apply do not 
shift the likelihood distributions outside the well-sampled regions.


\section{SUSY searches at the LHC}
\label{sec:susy}

The updates concerning SUSY searches in \citere{mc7} are
based on the public results of searches for supersymmetric 
signals using $\sim 1$/fb of LHC data analyzed by the ATLAS and CMS 
Collaborations and $\sim 0.3$/fb of data analyzed by the LHCb Collaboration.
For our purposes, some of the most important constraints are provided by the 
ATLAS~\cite{ATLASsusy} and CMS~\cite{CMSsusy} searches for jets +
$\ETslash$ events without leptons, as well as
searches for the heavier MSSM Higgs bosons, $H/A$~\cite{ATLASHA,CMSHA}. Also
important are the new upper limits on \bmm\ from the CMS~\cite{CMSbmm},
LHCb~\cite{LHCbbmm} and CDF Collaborations~\cite{CDFbmm}, 
which are incorporated in \citere{mc7}.

The CMS and ATLAS Collaborations have both announced new exclusions in the 
$(m_0, m_{1/2})$ plane of the CMSSM based on searches for events with
jets + $\ETslash$ unaccompanied by charged leptons, assuming $\tb = 10$, 
$A_0 = 0$ and $\mu > 0$. 
The updated CMS $\alpha_T$ analysis is based on 1.1/fb of
data~\cite{CMSsusy}, and the updated ATLAS 0-lepton analysis is based on
1.04/fb of data~\cite{ATLASsusy}. It is known that 0-lepton analyses are
in general relatively insensitive to the $\tb$ and $A_0$ 
parameters of the CMSSM, as has been confirmed specifically for the CMS
$\alpha_T$ analysis, and they are also insensitive to the amount of
Higgs non-universality in the NUHM1. Therefore, we treat these analyses
as constraints in the $(m_0, m_{1/2})$ planes of the CMSSM and NUHM1
that are independent of the other model parameters. 

The CMS and ATLAS 0-lepton searches are most powerful in complementary
regions of the $(m_0, m_{1/2})$ plane. Along each ray in this plane, we
compare the expected CMS and ATLAS sensitivities, select the search that
has the stronger expected 95\% CL limit, and apply the constraint
imposed by that search~%
\footnote{It would also facilitate the modelling of LHC constraints on 
supersymmetry if the results from different Collaborations were combined
officially, as was done at LEP, is already done for \bmm\ searches, and
is planned for Higgs searches.}%
.~This subsequent evaluation/application of additional/new $\chi^2$
contributions is called the `afterburner'. We assign $\Delta \chi^2 = 5.99$,
corresponding to 1.96 effective standard deviations, 
along the CMS and ATLAS 95\% 0-lepton exclusion contours in the 
$(m_0, m_{1/2})$ plane. In the absence of more complete experimental
information, we approximate the impact of these constraints by assuming
that event numbers scale along rays in this plane 
$\propto {\cal M}^{-4}$ where ${\cal M} \equiv \sqrt{m_0^2 + m_{1/2}^2}$, 
as described in \cite{mc6}. We then use these
numbers to calculate the effective numbers of standard deviations 
and corresponding values of $\Delta \chi^2$ at each point in the plane.  
This procedure has been validated by comparing the likelihood it yields
with results obtained independently using the generic detector
simulation code {\tt DELPHES}, which has also been shown to reproduce
quite accurately the likelihood function evaluated by the CMS
Collaboration using their data~\cite{JM}.


\section{Searches for a SM-like Higgs boson at the LHC}
\label{sec:higgs}

Within the supersymmetric frameworks discussed here, a confirmation of the
excess reported by ATLAS and CMS~\cite{Dec13} and consequently the discovery
of a SM-like Higgs boson is expected 
to be possible during this year, with a mass in the range 
between 114 and 130~GeV~\cite{Dec13}. We assume that this
measurement will yield a nominal value of $\Mh$ within this range, with
an experimental error that we estimate as $\pm 1\gev$. 
In \citere{mc75} the possibility that the LHC experiments confirm the
excess reported around $125 \gev$ and indeed discover a SM-like Higgs
boson was analyzed. Assuming
\begin{align}
\Mh = 125 \pm 1 ({\rm exp.}) \pm 1.5 ({\rm theo.}) \gev~,
\label{Mh125}
\end{align} 
this new constraint was incorporated using the `afterburner' approach 
discussed above (see also \citere{mc7}).


\section{Results from updated SUSY fits}
\label{sec:results}

\begin{figure*}[htb!]
\resizebox{7.5cm}{!}{\includegraphics{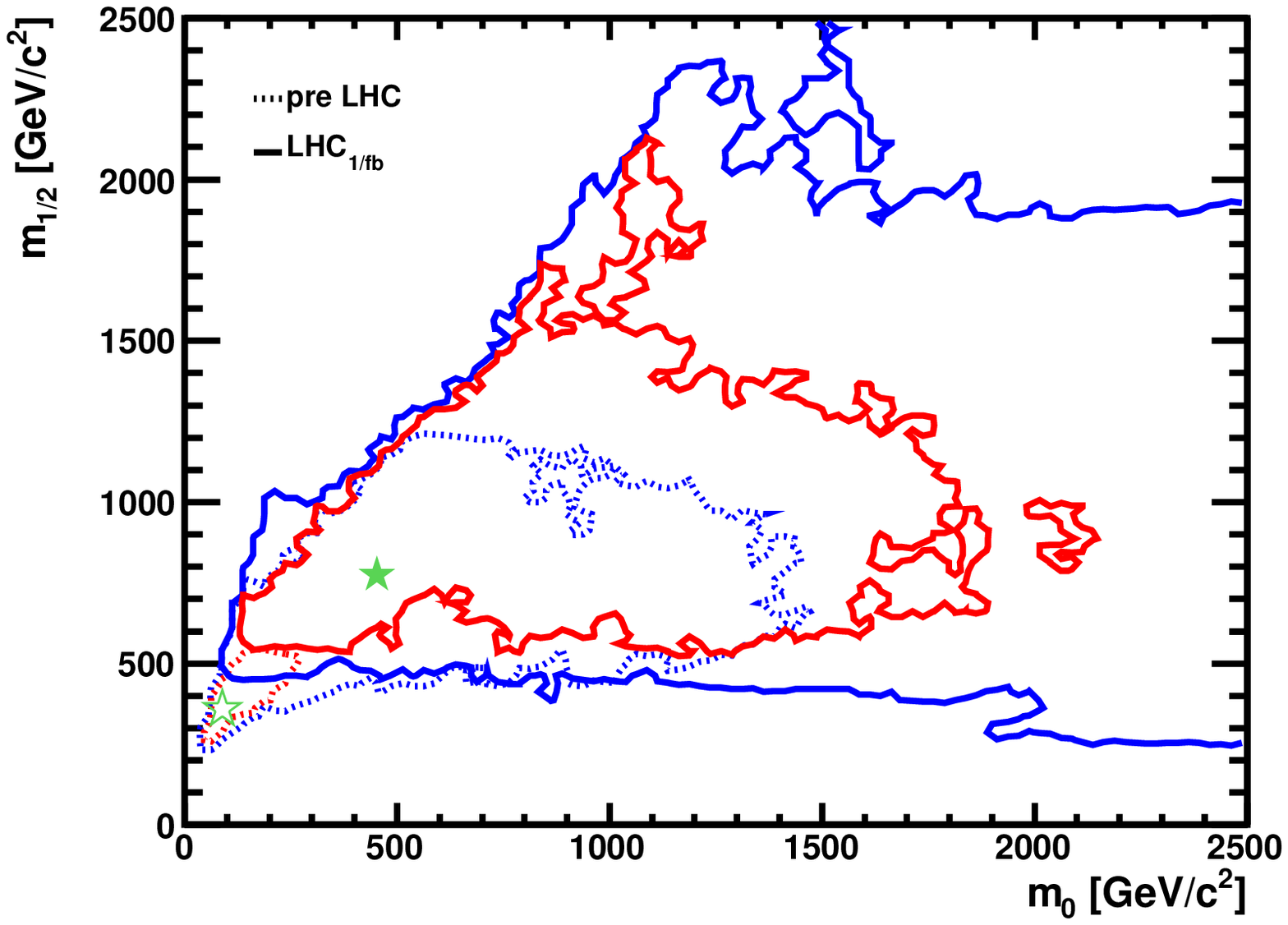}}
\resizebox{7.5cm}{!}{\includegraphics{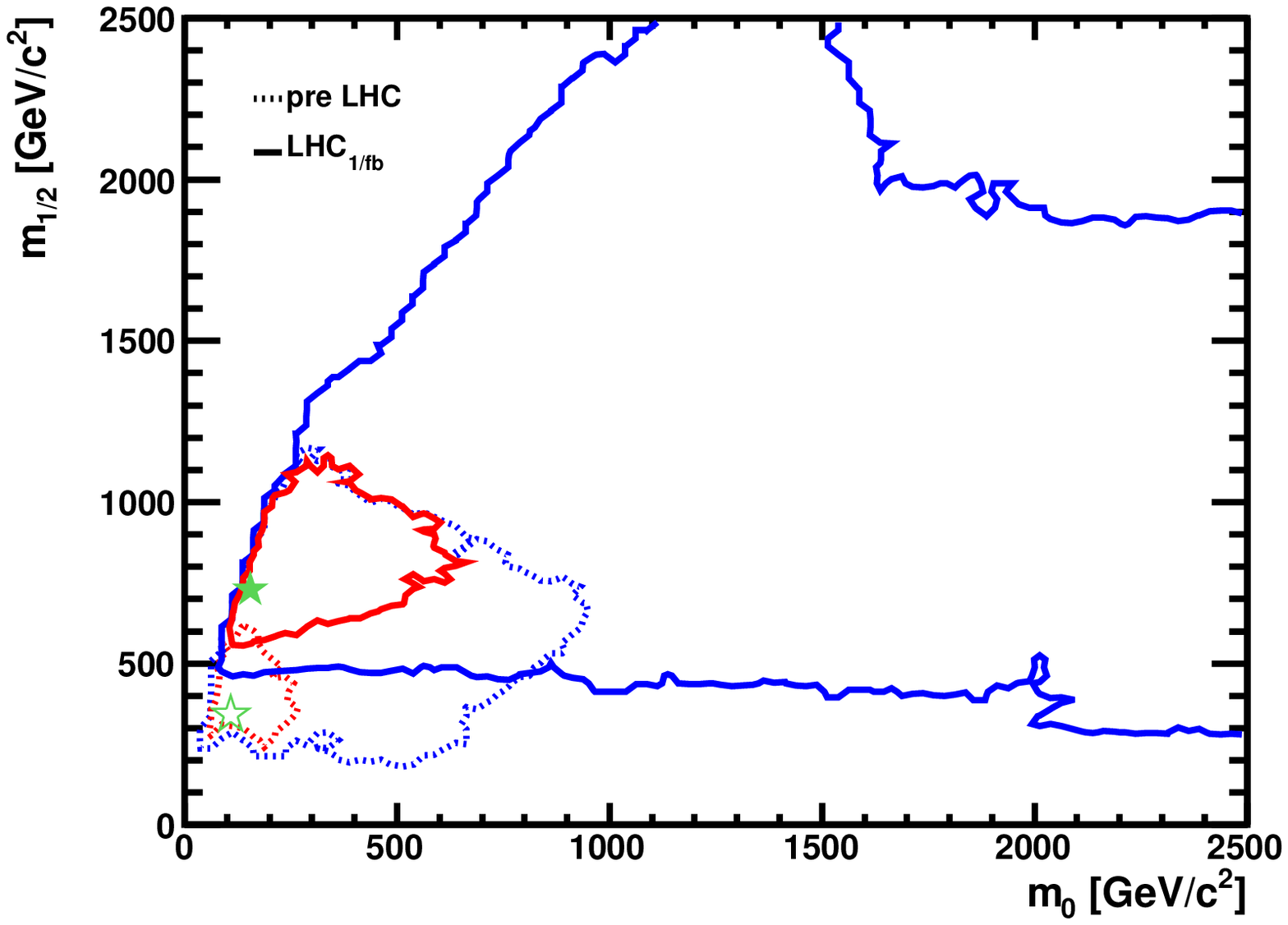}}\\
\resizebox{7.5cm}{!}{\includegraphics{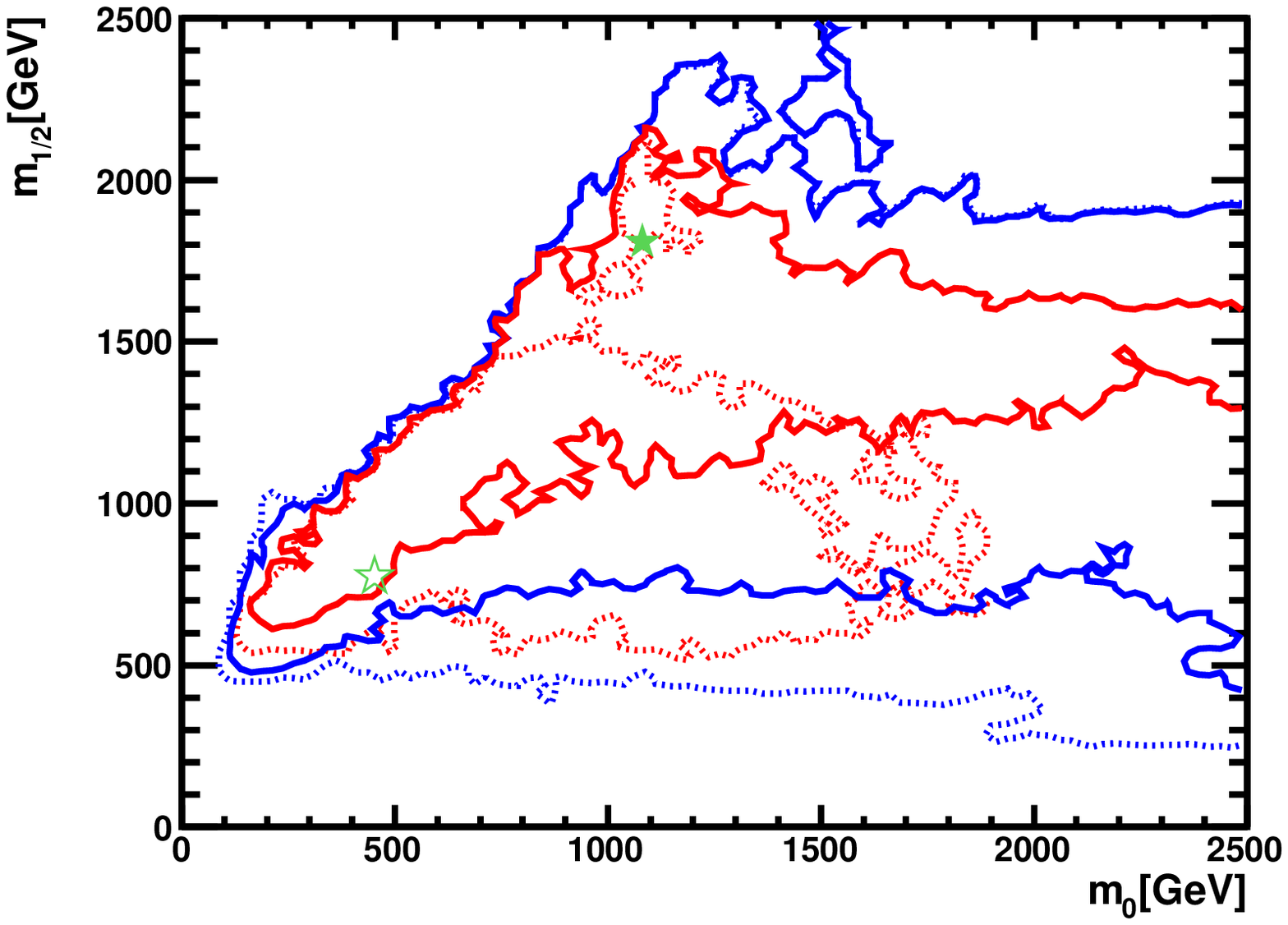}}
\resizebox{7.5cm}{!}{\includegraphics{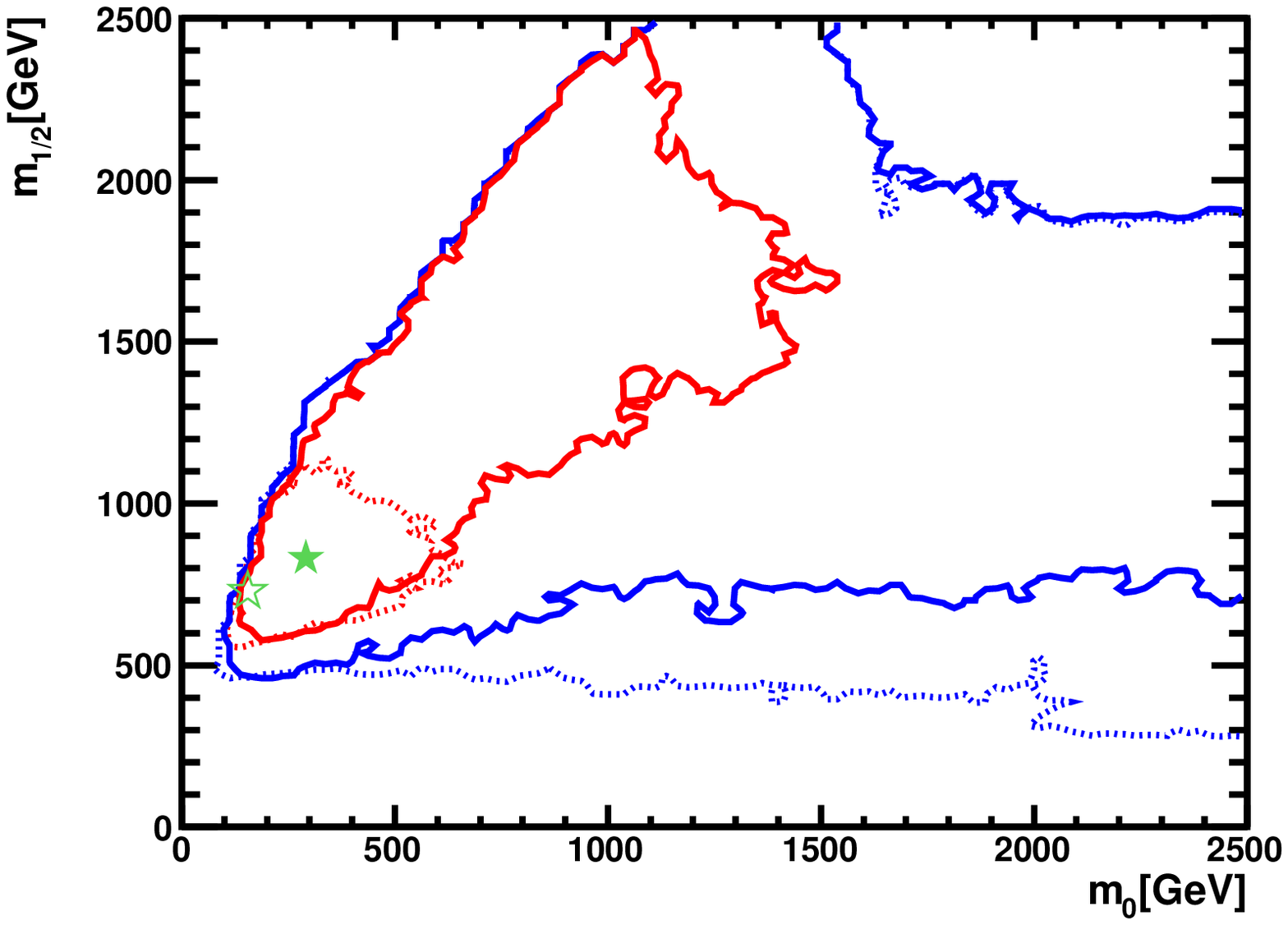}}
\caption{\it The $(m_0, m_{1/2})$ planes in the CMSSM (left) and the
  NUHM1 (right). The $\Delta \chi^2= 2.30$ and $5.99$   contours,
  commonly interpreted as the boundaries of the 68 and 95\% CL regions,
  are indicated in red and blue, respectively.
  In the upper row the best-fit point after incorporation of the
  LHC$_{\rm 1/fb}$ constraints is indicated by a filled green star, and the
  pre-LHC fit~\protect\cite{mc4} by an open star. The solid lines include the
  LHC$_{\rm 1/fb}$ data and the dotted lines showing the pre-LHC fits.
  In the lower row the solid lines include
  the hypothetical LHC measurement $\Mh = 125 \pm 1 \gev$
  and allowing for a theoretical error $\pm 1.5 \gev$,
  and the dotted lines repeat the contours including the LHC$_{\rm 1/fb}$,
  but without this $\Mh$ constraint.
  Here the open green stars denote the pre-Higgs best-fit
  points~\protect\cite{mc7}, whereas the solid green stars indicate the new
  best-fit points.  
}
\label{fig:6895}
\end{figure*}

We now review the effects on
the global likelihood functions in various CMSSM and NUHM1
parameter planes from the inclusion of the latest SUSY and Higgs search
data. Below also the implications for various physics observables are studied.
The $(m_0, m_{1/2})$ planes
shown in Fig.~\ref{fig:6895} are for the CMSSM (left) and NUHM1 (right).
The regions preferred at the 68\%~CL are outlined in red, and those
favoured at the 95\%~CL are outlined in blue. 
In the upper row the solid (dotted) lines
include (omit) the LHC$_{\rm 1/fb}$ data.
The open green star denotes the pre-LHC best-fit point~\cite{mc4},
whereas the solid green star indicates the new best-fit point incorporating
the LHC SUSY search results. 
In the lower row the solid (dotted) lines
include (omit) the assumed LHC 
Higgs constraint.
The open green star denotes the pre-Higgs best-fit point~\cite{mc7},
whereas the solid green star indicates the new best-fit point incorporating
a Higgs-boson mass measurement at $125 \gev$.

\begin{figure*}[htb!]
\resizebox{7.5cm}{!}{\includegraphics{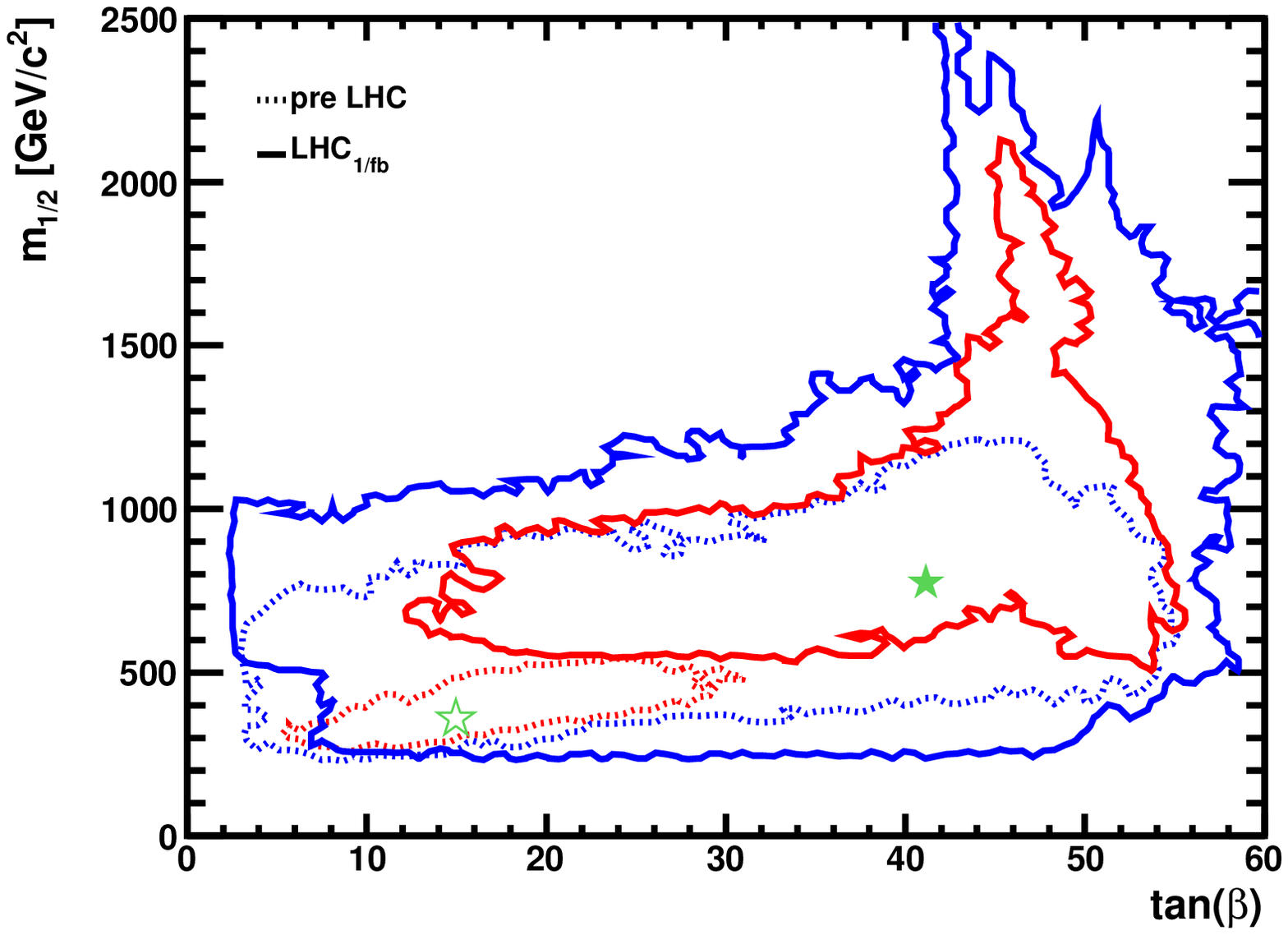}}
\resizebox{7.5cm}{!}{\includegraphics{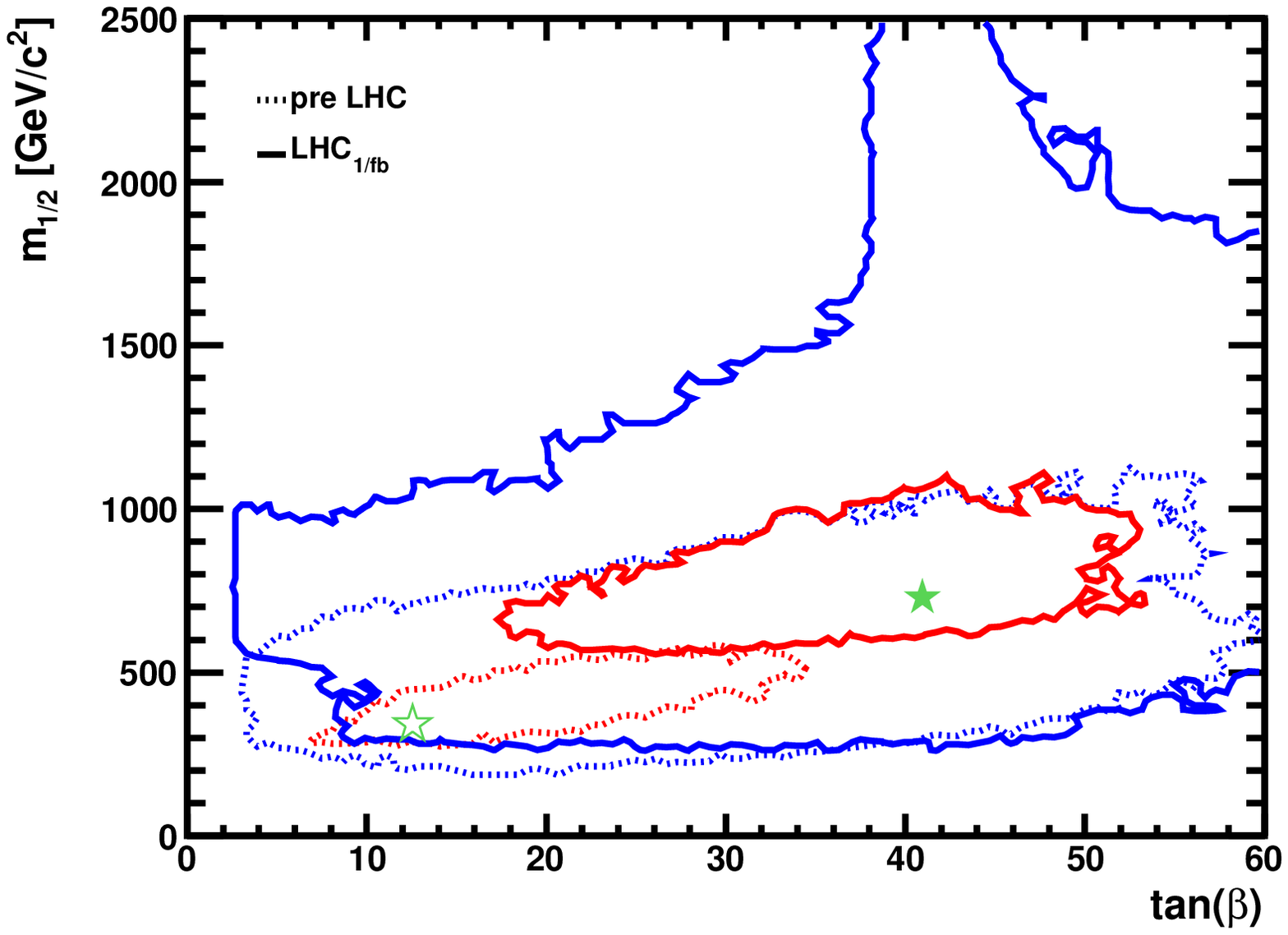}}\\
\resizebox{7.5cm}{!}{\includegraphics{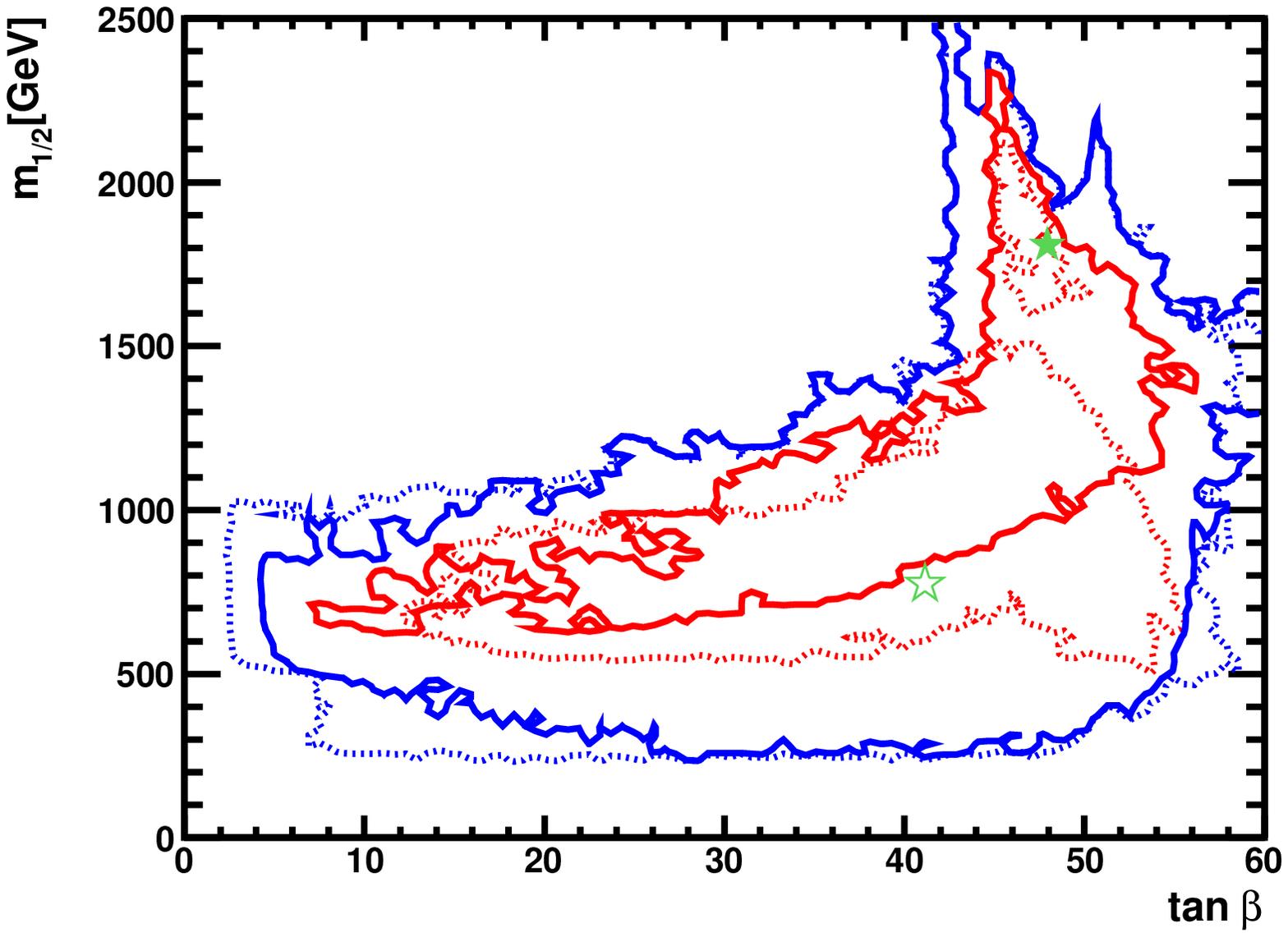}}
\resizebox{7.5cm}{!}{\includegraphics{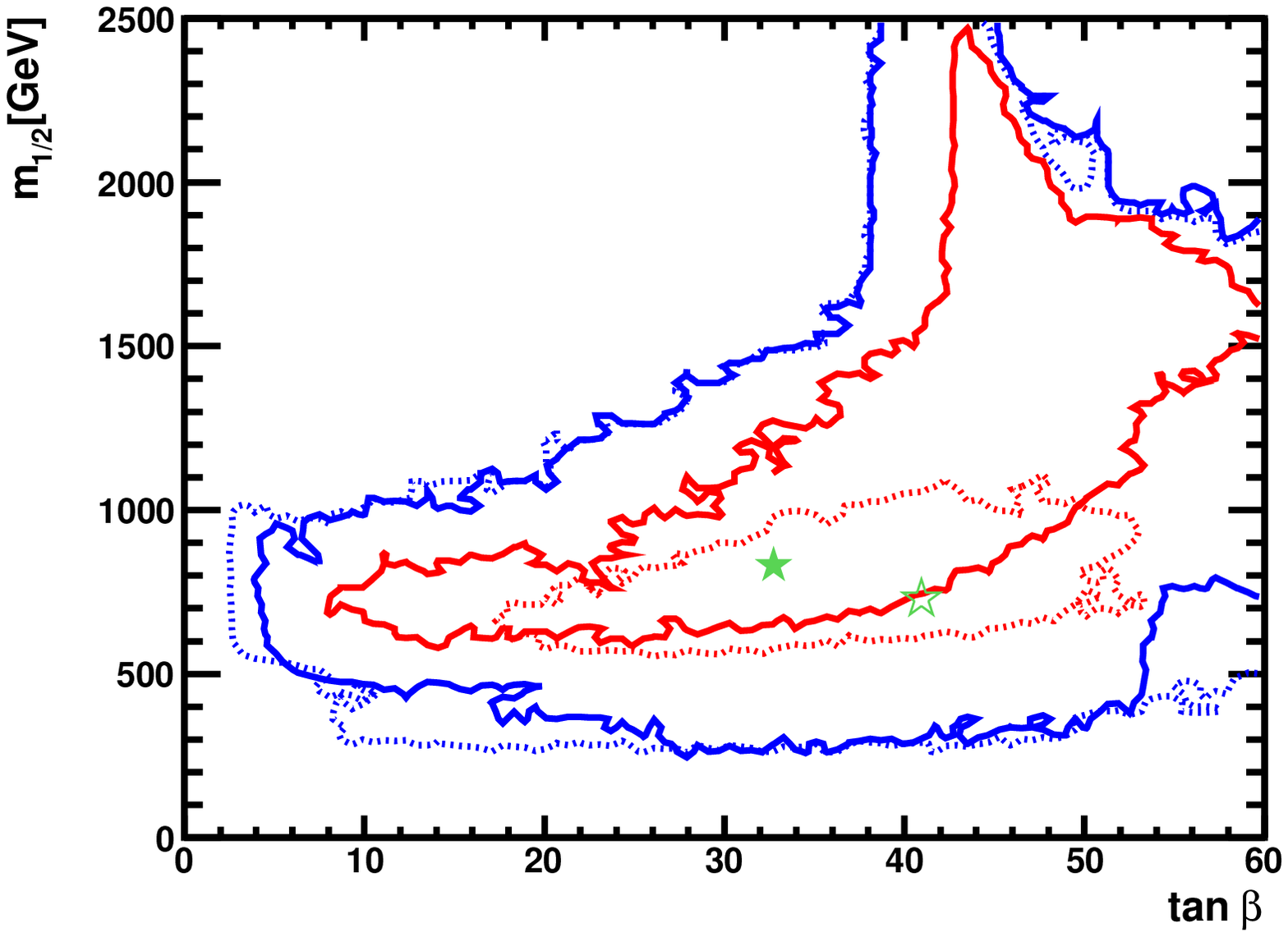}}
\vspace{-1cm}
\caption{\it The $(m_{1/2}, \tb)$ planes in the CMSSM (left) and the
  NUHM1 (right), including (omitting) the LHC$_{\rm 1/fb}$ constraints
  in the upper row, including (omitting) the the hypothetical LHC
  measurement $\Mh = 125 \pm 1 \gev$ in the lower row. The notations and
  significations of the contours are 
  the same as in Fig.~\protect\ref{fig:6895}.
}
\label{fig:tanbm12}
\end{figure*}

One can see in \reffi{fig:6895} that both new experimental results have
a similar, adding up effect. The preferred regions are shifted by both
new constraints to substantially
larger $m_{1/2}$ and somewhat larger $m_0$ values. A similar effect can
be observed in \reffi{fig:tanbm12}, where we show the $(m_{1/2}, \tb)$
planes in the CMSSM (left) and the NUHM1 (right), including (omitting)
the LHC$_{\rm 1/fb}$ constraints in the upper row, including (omitting)
the the hypothetical LHC measurement $\Mh = 125 \pm 1 \gev$ in the lower
row. The preferred values of $\tb$ are shifted to substantially higher
values.

\begin{figure*}[htb!]
\resizebox{7.5cm}{!}{\includegraphics{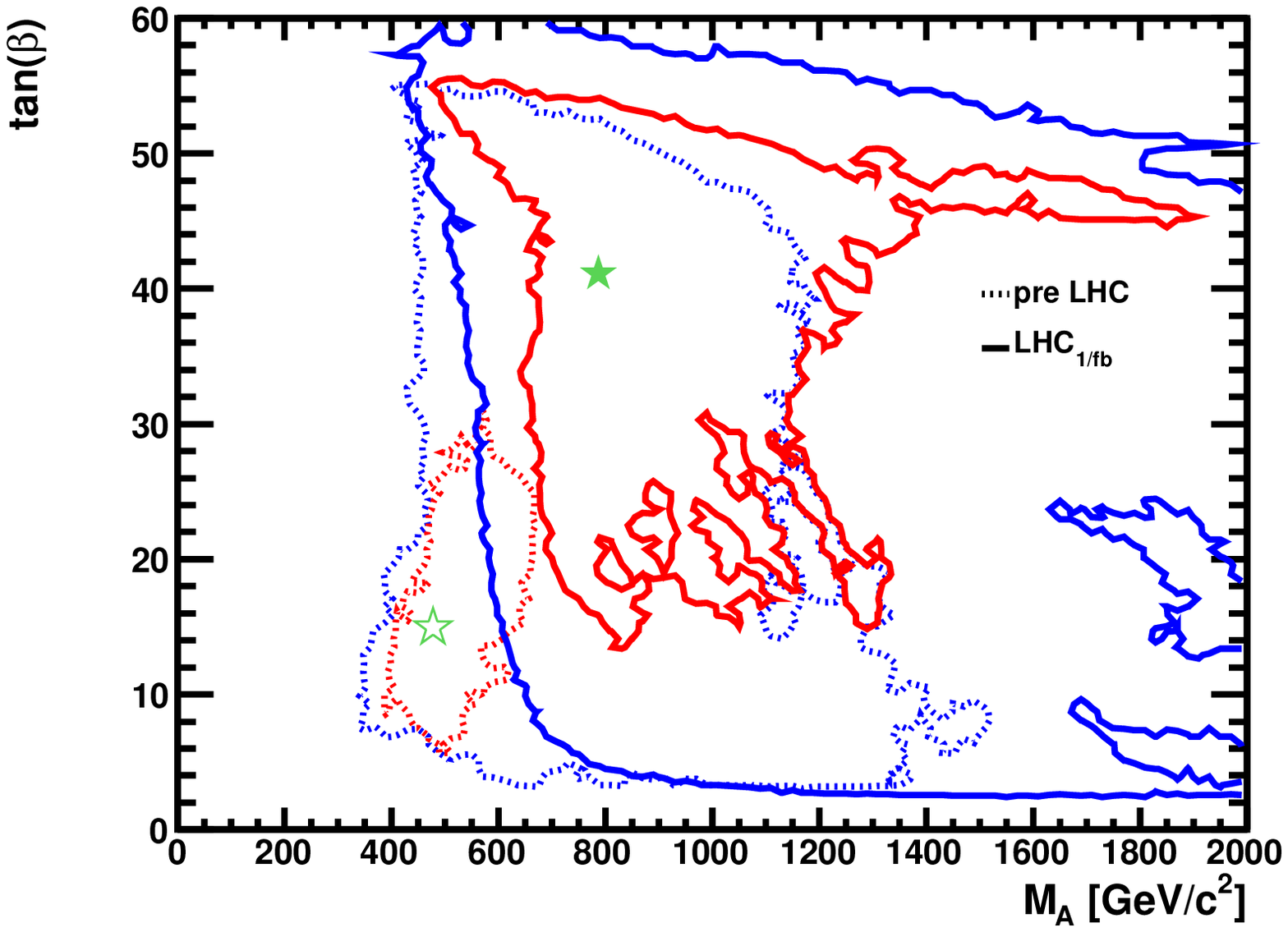}}
\resizebox{7.5cm}{!}{\includegraphics{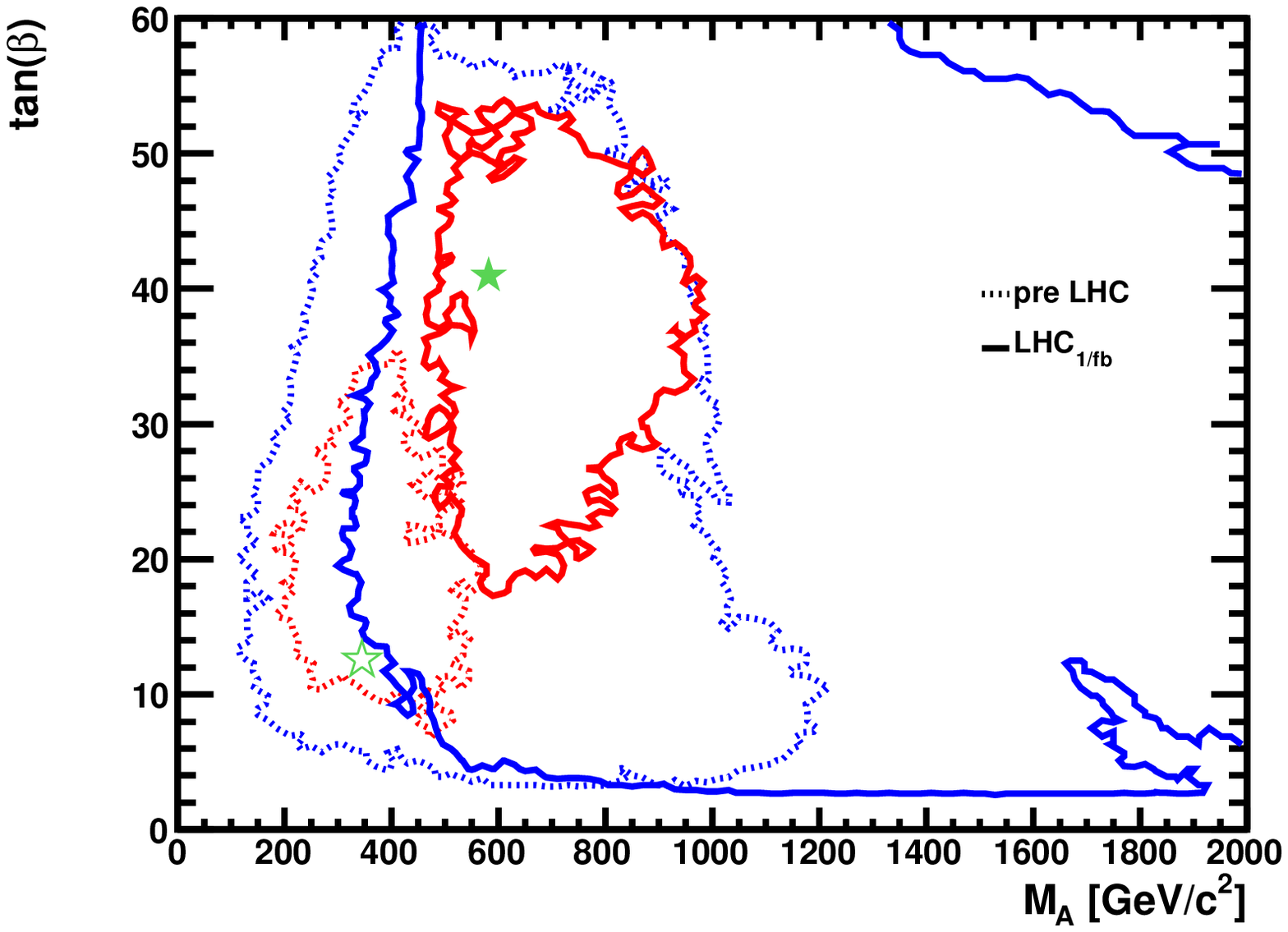}}\\
\resizebox{7.5cm}{!}{\includegraphics{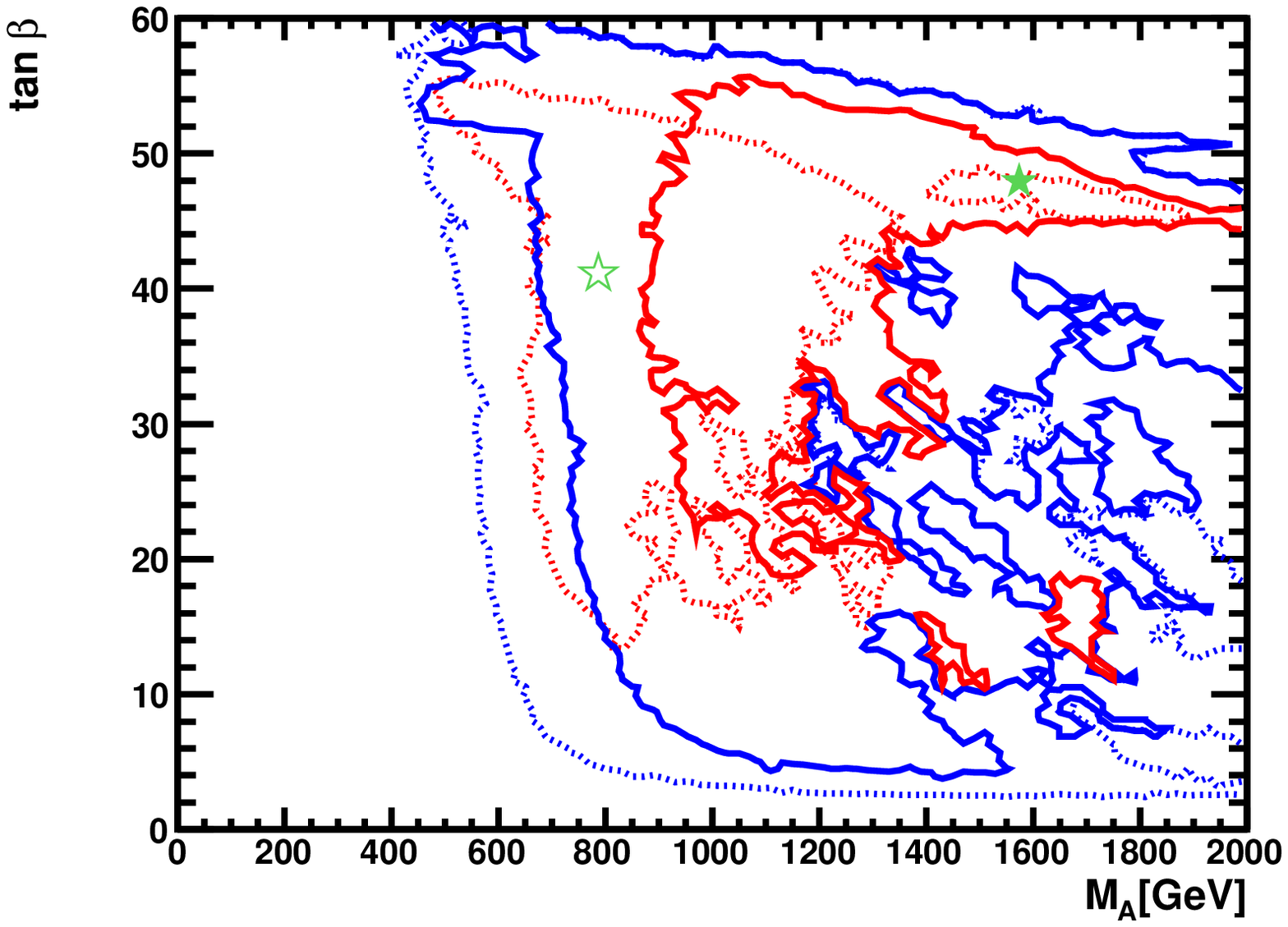}}
\resizebox{7.5cm}{!}{\includegraphics{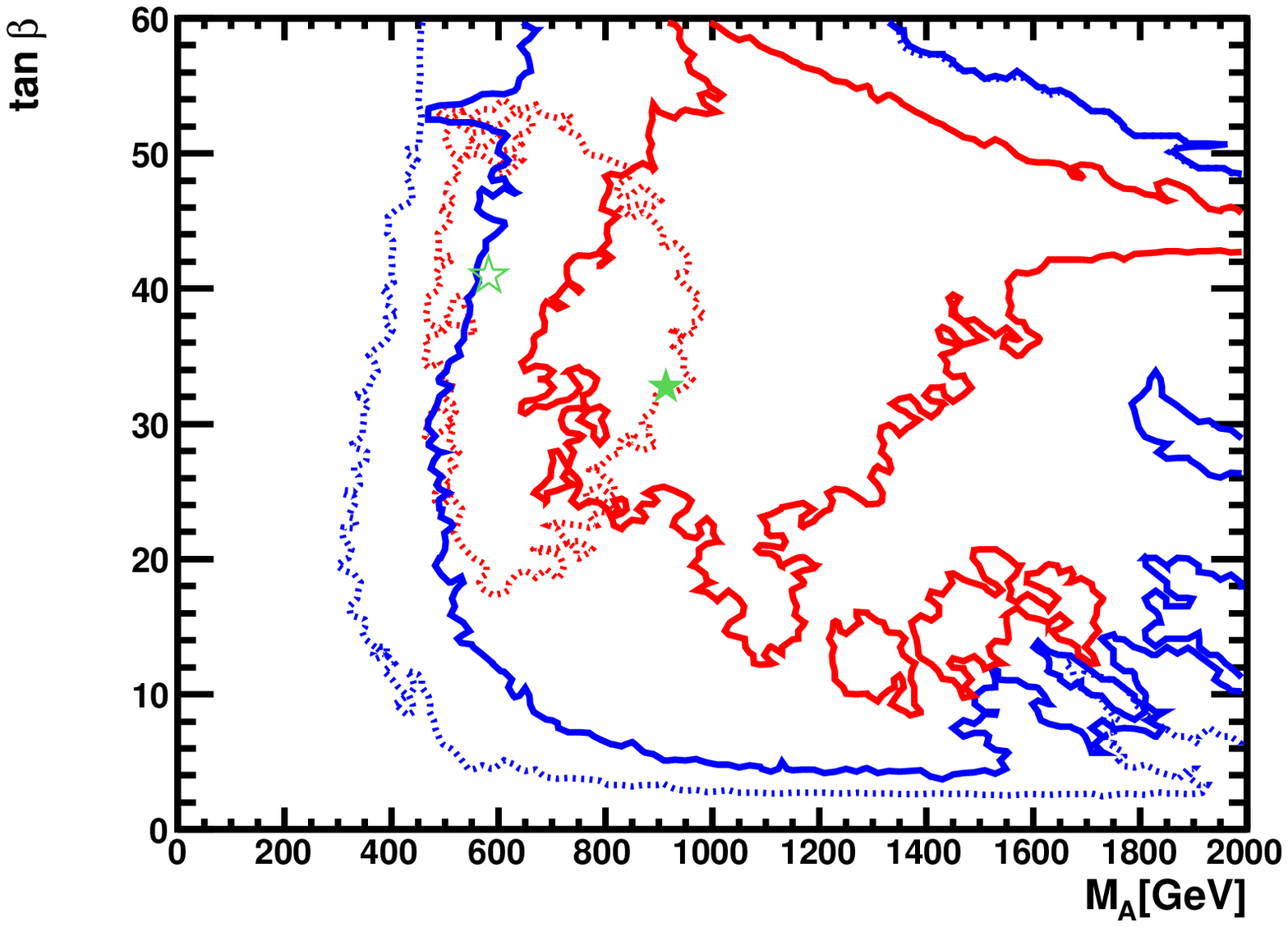}}
\vspace{-1cm}
\caption{\it The $(\MA, \tb)$ planes in the CMSSM (left) and the
  NUHM1 (right), including (omitting) the LHC$_{\rm 1/fb}$ constraints
  in the upper row, including (omitting) the the hypothetical LHC
  measurement $\Mh = 125 \pm 1 \gev$ in the lower row. The notations and
  significations of the contours are 
  the same as in Fig.~\protect\ref{fig:6895}.
}
\label{fig:MAtb}
\end{figure*}

The results for the $(\MA, \tb)$ planes in the CMSSM and the NUHM1
are shown in Fig.~\ref{fig:MAtb}. Again both new experimental results
show a similar effect. We observe a strong increase in the
best-fit value of $\MA$ in both models, especially in the CMSSM, where
after the inclusion of the latest SUSY searches and the hypothetical
Higgs signal now $\MA \sim 1600 \gev$ is preferred. 
However, the likelihood function varies relatively slowly in both
models, as compared to the pre-LHC fits. 

\section{Implications for the ILC(1000)}
\label{sec:ilc}

In view of the interest in building an $e^+ e^-$ collider as the next
major project at the energy frontier, we briefly review the
post-LHC$_{\rm 1/fb}$ predictions for expectations for sparticle
production in $e^+e^-$ annihilation within the CMSSM and NUHM1.
In this respect it has to be kept
in mind that the LHC searches are mainly sensitive to the 
production of coloured particles, whereas
lepton colliders will have a high sensitivity in particular for the 
production of colour-neutral states, such as
sleptons, charginos and neutralinos, as well as yielding high-precision
measurements that will provide indirect sensitivity to quantum effects of new
states. 

\begin{figure*}[htb!]
\begin{center}
\resizebox{6.5cm}{!}{\includegraphics{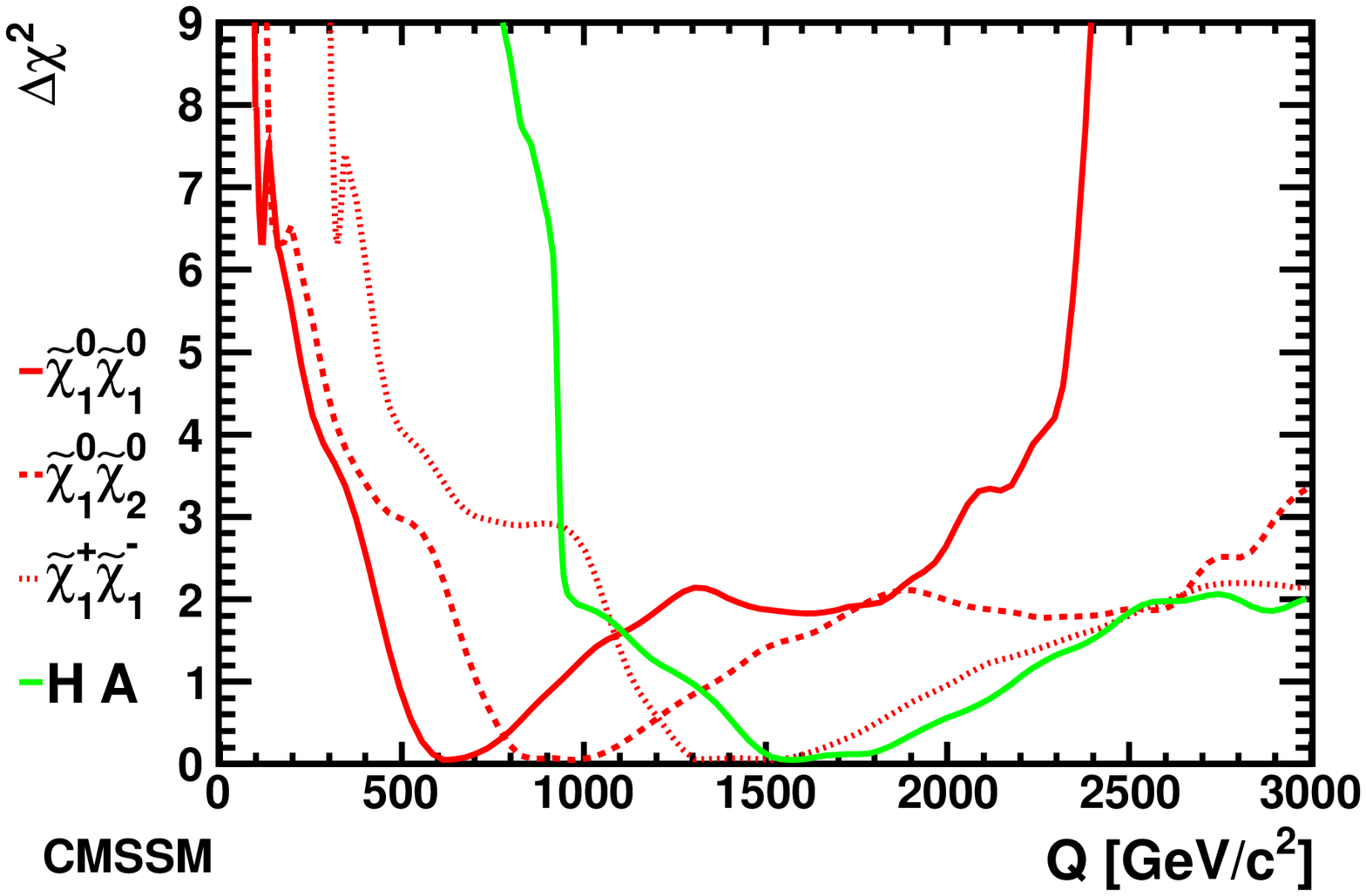}}
\resizebox{6.5cm}{!}{\includegraphics{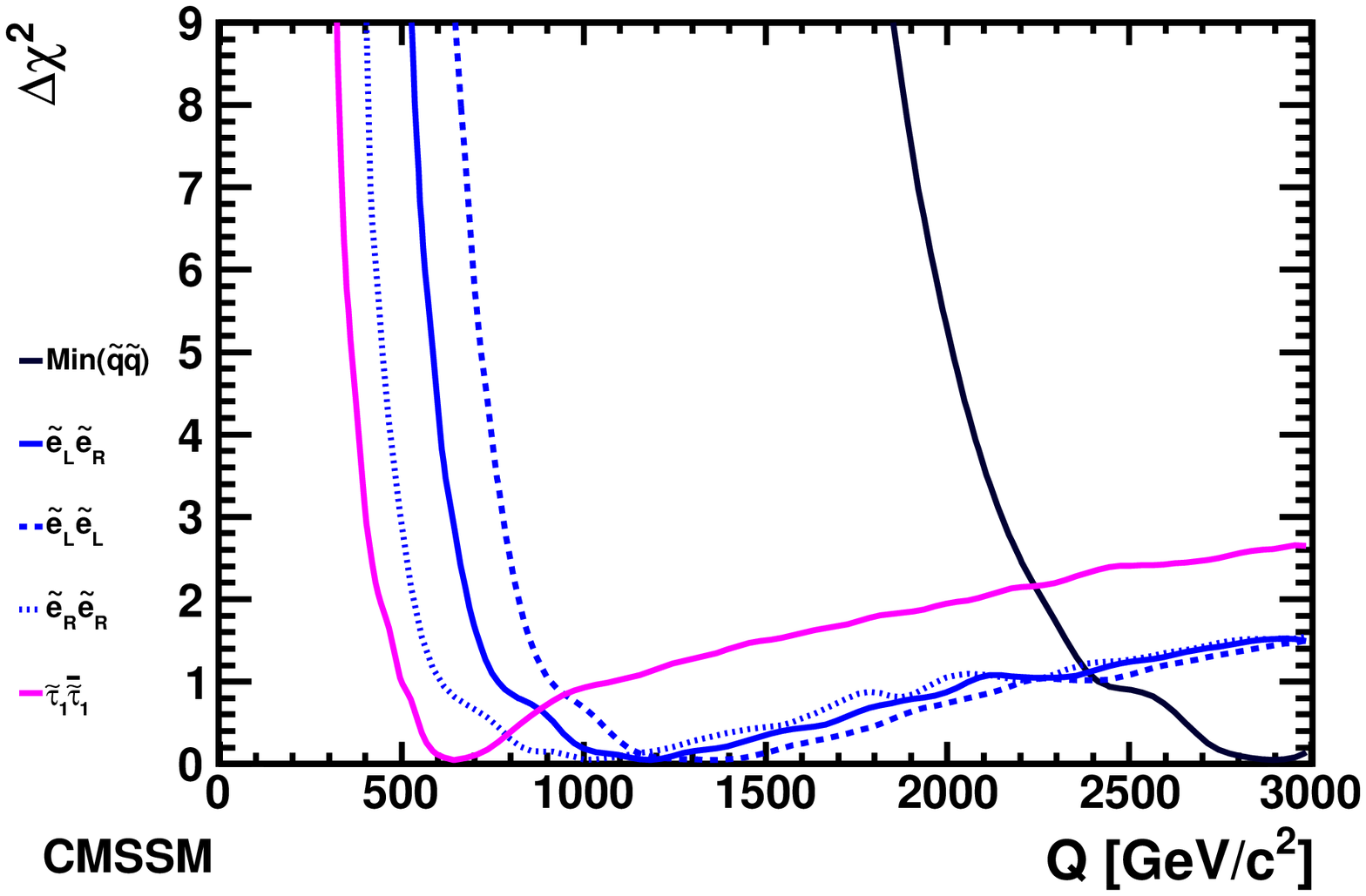}}\\
\resizebox{6.5cm}{!}{\includegraphics{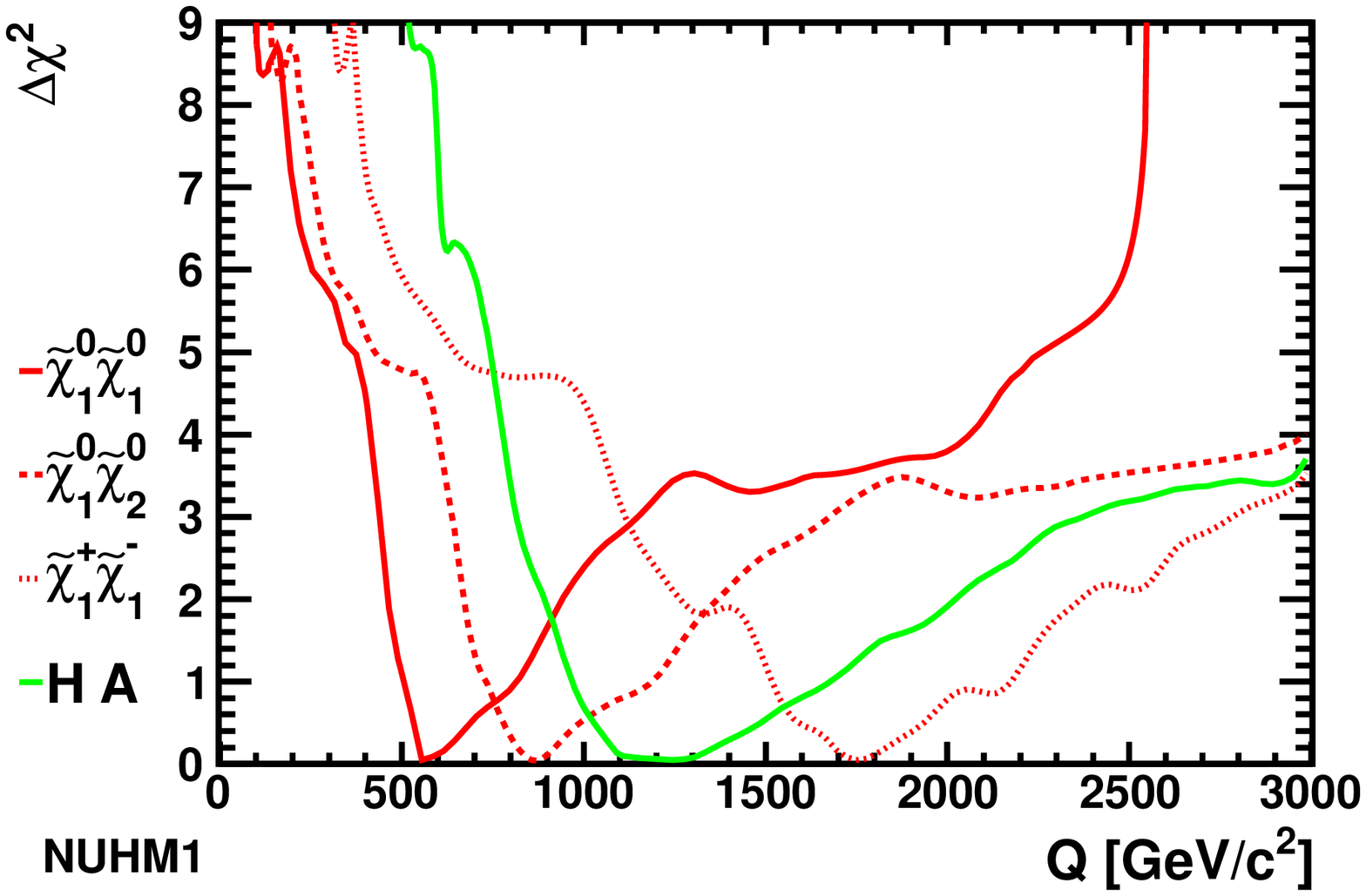}}
\resizebox{6.5cm}{!}{\includegraphics{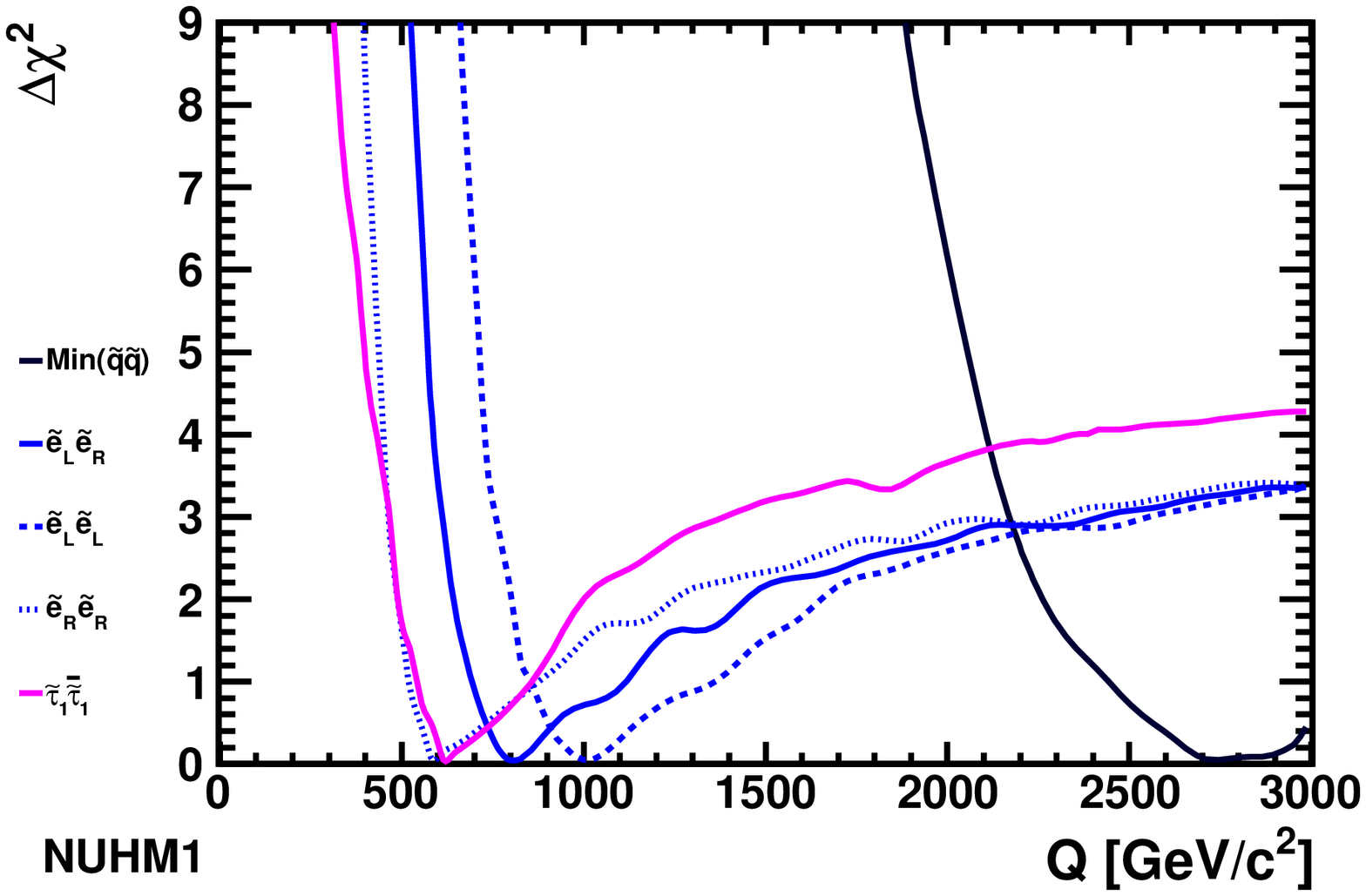}}\\
\vspace{-1cm}
\end{center}
\caption{\it The $\chi^2$ likelihood functions for various pair-production
thresholds in $e^+ e^-$, as estimated in the CMSSM (upper panel) and the NUHM1 (lower panel)
after incorporating the XENON100~\protect\cite{XE100} and LHC$_{\rm 1/fb}$ constraints.
The likelihood function for the $\asmu{R}\smu{R}$ threshold (not shown)
is very similar to that for $\asel{R}\sel{R}$.
}
\label{fig:thresholds}
\end{figure*}

Fig.~\ref{fig:thresholds} compares the likelihood functions for
various thresholds in the CMSSM (upper panel) and the NUHM1 (lower panel),
based on the global fits made using the LHC$_{\rm 1/fb}$ and XENON100
constraints. The lowest thresholds are those for
$e^+ e^- \to \neu{1} \neu{1}$, $\astaue \staue$, $\asel{R}\sel{R}$
 and $\asmu{R}\smu{R}$ (the latter is not shown, it is similar
to that for $\asel{R}\sel{R}$). We see that, within the CMSSM and 
NUHM1, it now seems that these thresholds may well lie above 500~GeV,
though in the CMSSM significant fractions of their 
likelihood functions still lie below 500~GeV.
The thresholds for $\neu{1} \neu{2}$ and
$\asel{R}\sel{L} + \asel{L}\sel{R}$ are expected to be somewhat higher,
possibly a bit below 1~TeV.
The preferred value for the threshold for $\cha{1}\champ{1}$ 
lies at about 1700 GeV in both the CMSSM and NUHM1 scenarios, 
that for the $HA$ threshold lies above 1~TeV, and that
for first- and second-generation squark-antisquark pair production lies
beyond 2.5~TeV in both models. 

\begin{table*}[htb!]
\renewcommand{\arraystretch}{1.5}
\begin{center}
\begin{tabular}{|c||c|c|c|c|c|c|} \hline
Model & Minimum & Fit Prob- & $m_{1/2}$ & $m_0$ & $A_0$ & $\tb$ \\
      & $\chi^2$/d.o.f.& ability & (GeV) & (GeV) & (GeV) & \\
\hline \hline
CMSSM  & & & & & &  \\
\hline
pre-LHC 
    & 21.5/20 & {37\%} & $360_{-100}^{+180}$ & $90_{-50}^{+220}$ 
    & $-400^{+730}_{-970}$ & $15_{-9}^{+15}$ \\
pre-Higgs, post-LHC
                     & 28.8/22 & 15\% & 780 & 450 & $-1110$ & 41 \\
$\Mh \simeq 125 \gev$ & 30.6/23 & 13\% & 1800 & 1080 & $ 860$ & 48 \\
\hline\hline
NUHM1 & & & & & &  \\
\hline
pre-LHC    
    & 20.8/18 & 29\% & $340_{-110}^{+280}$ & $110_{-30}^{+160}$ 
    & $520^{+750}_{-1730}$ & $13_{-6}^{+27}$ \\
pre-Higgs, post-LHC
                     & 26.9/21 & 17\% &  730 &  150  & $-910$ & 41 \\
$\Mh \simeq 125 \gev$ & 29.7/22 & 13\% & 830 & 290 & $ 660$ & 33 \\
\hline
\end{tabular}
\caption{\it Comparison of the best-fit points found in the CMSSM and NUHM1
  pre-LHC$_{\rm 1/fb}$~\cite{mc4}, pre-Higgs~\cite{mc7} and including
  the latest results from SUSY and Higgs searches.
}
\label{tab:p-comp}
\end{center}
\end{table*}

\subsection*{What are the implications for the ILC(1000)?}

Two aspects are important for the correct interpretation of the results
reviewed above with respect to the ILC(1000).
First, anything inferred from the coloured sector concerning the
uncoloured sector depends on the underlying model assumptions, and in
particular on  assumptions about the possible universality of soft
supersymmetry breaking at the 
GUT scale. Non-universal models, e.g., low-energy supersymmetric
models, or models with different GUT assumptions,
could present very different possibilities.

Second, while within the CMSSM and the NUHM1 the preferred SUSY and
heavy Higgs mass ranges move to substantially higher values,
this upward shift goes along with a substantial increase in
$\chi^2$/d.o.f., as can be seen in \refta{tab:p-comp}. Here we review
the results for the best-fit points in the CMSSM and in the NUHM1
``pre-LHC'' (i.e.\ no SUSY searches at the LHC, nor any assumption about
about any possible Higgs mass measurement), ``pre-Higgs, post-LHC''
(i.e.\ including the LHC$_{\rm 1/fb}$ SUSY searches, but no assumption
about any possible Higgs measurement) and ``$\Mh \simeq 125 \gev$'' (in
addition the assumption of \refeq{Mh125}). The droop in the ``Fit
probability'' is clearly visible. The reason is that the pre-LHC data
favors relatively low SUSY mass scales (to a large extent driven
by \gmt, supported by $\MW$ and other observables), while the
non-observation of colored SUSY particles favores a heavier spectrum,
leading to an increasing tension within the CMSSM and the NUHM1.
Concerning the production thresholds shown in \reffi{fig:thresholds}, 
further increases in the
excluded regions would yield even higher thresholds, but would also
make the  CMSSM or NUHM1 seem even less likely.
The time might come to take another look at other GUT based or 
non-minimal supersymmetric models, in which the colored sector (mainly
searched for at the LHC) and the uncolored sector (which is relevant,
e.g., for \gmt) are less strongly connected as in the CMSSM or NUHM1.


\section*{Acknowledgments}

We thank 
O.~Buchmueller, 
R.~Cavanaugh, 
A.~De Roeck, 
M.J.~Dolan, 
J.R.~Ellis, 
H.~Fl\"acher, 
G.~Isidori,
J.~Marrouche,
K.A.~Olive,
S.~Rogerson,
F.J.~Ronga,
K.J.~de~Vries
and
G.~Weiglein
with whom many results shown here have been obtained.
We thank especially G.~Moortgat-Pick for helpful discussions and
important contributions to the talk.
The work of S.H. is supported 
in part by CICYT (grant FPA 2010--22163-C02-01) and by the
Spanish MICINN's Consolider-Ingenio 2010 Program under grant MultiDark
CSD2009-00064.



\begin{footnotesize}


\end{footnotesize}


\begin{thebibliography}{99}

%
%

\bibitem{mc1}
O.~Buchmueller {\it et al.},
  Phys.\ Lett.\  B {\bf 657} (2007) 87
  [arXiv:0707.3447 [hep-ph]].

\bibitem{mc2}
  O.~Buchmueller {\it et al.},
  JHEP {\bf 0809} (2008) 117
  [arXiv:0808.4128 [hep-ph]].

\bibitem{mc3}
  O.~Buchmueller {\it et al.},
  Eur.\ Phys.\ J.\  C {\bf 64} (2009) 391
  [arXiv:0907.5568 [hep-ph]].

\bibitem{mc35}
  O.~Buchmueller {\it et al.},
  Phys.\ Rev.\  D {\bf 81} (2010) 035009
  [arXiv:0912.1036 [hep-ph]].

\bibitem{mc4}
  O.~Buchmueller {\it et al.},
  Eur.\ Phys.\ J.\  C {\bf 71} (2011) 1583
  [arXiv:1011.6118 [hep-ph]].

\bibitem{mc5}
  O.~Buchmueller {\it et al.},
  Eur.\ Phys.\ J.\  C {\bf 71} (2011) 1634
  [arXiv:1102.4585 [hep-ph]].

\bibitem{mc6}
O.~Buchmueller {\it et al.},
  Eur.\ Phys.\ J.\  C {\bf 71} (2011) 1722
  [arXiv:1106.2529 [hep-ph]].

\bibitem{mc7}
O.~Buchmueller {\it et al.},
to appear in Eur.\ Phys. J.\ C, 
  arXiv:1110.3568 [hep-ph].

\bibitem{mc75} O.~Buchmueller {\it et al.},
  arXiv:1112.3564 [hep-ph].

\bibitem{mc-web}
For more information and updates, please see {\tt http://cern.ch/mastercode/}.

\bibitem{pre-LHC}
For a sampling of other pre-LHC analyses, see:
  E.~A.~Baltz and P.~Gondolo,
  JHEP {\bf 0410} (2004) 052
  [arXiv:hep-ph/0407039];
  B.~C.~Allanach and C.~G.~Lester,
  Phys.\ Rev.\  D {\bf 73} (2006) 015013
  [arXiv:hep-ph/0507283];
  R.~R.~de Austri, R.~Trotta and L.~Roszkowski,
  JHEP {\bf 0605} (2006) 002
  [arXiv:hep-ph/0602028];
  R.~Lafaye, T.~Plehn, M.~Rauch and D.~Zerwas,
  Eur.\ Phys.\ J.\  C {\bf 54} (2008) 617
  [arXiv:0709.3985 [hep-ph]];
  S.~Heinemeyer, X.~Miao, S.~Su and G.~Weiglein,
  JHEP {\bf 0808} (2008) 087
  [arXiv:0805.2359 [hep-ph]];
  R.~Trotta, F.~Feroz, M.~P.~Hobson, L.~Roszkowski and R.~Ruiz de Austri,
  JHEP {\bf 0812} (2008) 024
  [arXiv:0809.3792 [hep-ph]];
  P.~Bechtle, K.~Desch, M.~Uhlenbrock and P.~Wienemann,
  Eur.\ Phys.\ J.\  C {\bf 66} (2010) 215
  [arXiv:0907.2589 [hep-ph]].

\bibitem{post-LHC} 
  For a sampling of other post-LHC analyses, see:
 D.~Feldman, K.~Freese, P.~Nath, B.~D.~Nelson and G.~Peim,
  Phys.\ Rev.\ D {\bf 84}, 015007 (2011)
  [arXiv:1102.2548 [hep-ph]];
   B.~C.~Allanach,
  Phys.\ Rev.\ D {\bf 83}, 095019 (2011)
  [arXiv:1102.3149 [hep-ph]];
 S.~Scopel, S.~Choi, N.~Fornengo and A.~Bottino,
  Phys.\ Rev.\ D {\bf 83}, 095016 (2011)
  [arXiv:1102.4033 [hep-ph]];
  P.~Bechtle {\it et al.},
  arXiv:1102.4693 [hep-ph];
 B.~C.~Allanach, T.~J.~Khoo, C.~G.~Lester and S.~L.~Williams,
  JHEP {\bf 1106}, 035 (2011)
  [arXiv:1103.0969 [hep-ph]];
   S.~Akula, N.~Chen, D.~Feldman, M.~Liu, Z.~Liu, P.~Nath and G.~Peim,
  Phys.\ Lett.\ B {\bf 699}, 377 (2011)
  [arXiv:1103.1197 [hep-ph]];
M.~J.~Dolan, D.~Grellscheid, J.~Jaeckel, V.~V.~Khoze and P.~Richardson,
  JHEP {\bf 1106}, 095 (2011)
  [arXiv:1104.0585 [hep-ph]];
  S.~Akula, D.~Feldman, Z.~Liu, P.~Nath and G.~Peim,
  Mod.\ Phys.\ Lett.\ A {\bf 26}, 1521 (2011)
  [arXiv:1103.5061 [hep-ph]];
  M.~Farina, M.~Kadastik, D.~Pappadopulo, J.~Pata, M.~Raidal and A.~Strumia,
  Nucl.\ Phys.\ B {\bf 853}, 607 (2011)
  [arXiv:1104.3572 [hep-ph]];
S.~Profumo,
  Phys.\ Rev.\ D {\bf 84}, 015008 (2011)
  [arXiv:1105.5162 [hep-ph]];
    T.~Li, J.~A.~Maxin, D.~V.~Nanopoulos and J.~W.~Walker,
  arXiv:1106.1165 [hep-ph];
 N.~Bhattacharyya, A.~Choudhury and A.~Datta,
  Phys.\ Rev.\ D {\bf 84}, 095006 (2011)
  [arXiv:1107.1997 [hep-ph]].

  \bibitem{HK}
  H.~P.~Nilles, Phys. Rep. {\bf 110} (1984) 1;
  H.~E.~Haber and G.~L.~Kane,
  Phys.\ Rept.\  {\bf 117} (1985) 75.

\bibitem{EHNOS} H.~Goldberg,
                Phys.\ Rev.\ Lett.\ {\bf 50} (1983) 1419;
                J.~Ellis, J.~Hagelin, D.~Nanopoulos, K.~Olive and M.~Srednicki,
                Nucl.\ Phys.\ B {\bf 238} (1984) 453.

\bibitem{Komatsu:2010fb}
  E.~Komatsu {\it et al.}  [WMAP Collaboration],
  Astrophys.\ J.\ Suppl.\  {\bf 192} (2011) 18
  [arXiv:1001.4538 [astro-ph.CO]]; 
  {\tt http://lambda.gsfc.nasa.gov/product/map/current/parameters.cfm}.

\bibitem{ATLASsusy}
G.~Aad {\it et al.}  [ATLAS Collaboration],
  arXiv:1109.6572 [hep-ex].

\bibitem{ATLASHA}
ATLAS Collaboration, 
{\tt https://atlas.web.cern.ch/Atlas/GROUPS/PHYSICS/CONFNOTES/}\\
{\tt ATLAS-CONF-2011-132/ATLAS-CONF-2011-132.pdf}.

\bibitem{CMSsusy}
S.~Chatrchyan {\it et al.}  [CMS Collaboration],
  arXiv:1109.2352 [hep-ex].

\bibitem{CMSHA}
CMS Collaboration, 
{\tt http://cdsweb.cern.ch/record/1378096/files/HIG-11-020-pas.pdf.}

\bibitem{CMSbmm}
S.~Chatrchyan {\it et al.}  [CMS Collaboration],
  arXiv:1107.5834 [hep-ex].

\bibitem{LHCbbmm}
R.~Aaij {\it et al.}  [LHCb Collaboration],
  Phys.\ Lett.\  B {\bf 699} (2011) 330
  [arXiv:1103.2465 [hep-ex]];
  arXiv:1112.1600 [hep-ex].

\bibitem{Dec13}
F.~Gianotti for the ATLAS Collaboration, G.~Tonelli for the CMS Collaboration,\\
{\tt http://indico.cern.ch/conferenceDisplay.py?confId=164890}.

  \bibitem{newDavier}
M.~Davier, A.~Hoecker, B.~Malaescu and Z.~Zhang,
  Eur.\ Phys.\ J.\  C {\bf 71} (2011) 1515
  [arXiv:1010.4180 [hep-ph]].

 \bibitem{Jegerlehner}
F.~Jegerlehner and R.~Szafron,
  Eur.\ Phys.\ J.\  C {\bf 71} (2011) 1632
  [arXiv:1101.2872 [hep-ph]].

\bibitem{Allanach:2001kg}
  B.~C.~Allanach,
  Comput.\ Phys.\ Commun.\  {\bf 143} (2002) 305
  [arXiv:hep-ph/0104145].

  \bibitem{FeynHiggs}
 G.~Degrassi, S.~Heinemeyer, W.~Hollik, P.~Slavich and G.~Weiglein,
  Eur.\ Phys.\ J.\ C {\bf 28} (2003) 133
  [arXiv:hep-ph/0212020];
   S.~Heinemeyer, W.~Hollik and G.~Weiglein,
  Eur.\ Phys.\ J.\ C {\bf 9} (1999) 343
  [arXiv:hep-ph/9812472];
  S.~Heinemeyer, W.~Hollik and G.~Weiglein,
  Comput.\ Phys.\ Commun.\  {\bf 124} (2000) 76
  [arXiv:hep-ph/9812320];
   M.~Frank {\it et al.}, 
  JHEP {\bf 0702} (2007) 047
  [arXiv:hep-ph/0611326];
  See {\tt http://www.feynhiggs.de}~.

\bibitem{Moroi:1995yh}
  T.~Moroi,
  Phys.\ Rev.\  D {\bf 53} (1996) 6565
  [Erratum-ibid.\  D {\bf 56} (1997) 4424]
  [arXiv:hep-ph/9512396].

\bibitem{Degrassi:1998es}
  G.~Degrassi and G.~F.~Giudice,
  Phys.\ Rev.\  D {\bf 58} (1998) 053007
  [arXiv:hep-ph/9803384].

\bibitem{Heinemeyer:2003dq}
  S.~Heinemeyer, D.~Stockinger and G.~Weiglein,
  Nucl.\ Phys.\  B {\bf 690} (2004) 62
  [arXiv:hep-ph/0312264].

\bibitem{Heinemeyer:2004yq}
  S.~Heinemeyer, D.~Stockinger and G.~Weiglein,
  Nucl.\ Phys.\  B {\bf 699} (2004) 103
  [arXiv:hep-ph/0405255].

\bibitem{SuFla}
 G.~Isidori and P.~Paradisi,
  Phys.\ Lett.\ B {\bf 639} (2006) 499
  [arXiv:hep-ph/0605012];
  G.~Isidori, F.~Mescia, P.~Paradisi and D.~Temes,
  Phys.\ Rev.\  D {\bf 75} (2007) 115019
  [arXiv:hep-ph/0703035], and references therein.

\bibitem{SuperIso}
F.~Mahmoudi,
  Comput.\ Phys.\ Commun.\  {\bf 178} (2008) 745
  [arXiv:0710.2067 [hep-ph]]; 
  Comput.\ Phys.\ Commun.\  {\bf 180} (2009) 1579
  [arXiv:0808.3144 [hep-ph]];
  D.~Eriksson, F.~Mahmoudi and O.~Stal,
  JHEP {\bf 0811} (2008) 035
  [arXiv:0808.3551 [hep-ph]].

\bibitem{Svenetal}
  S.~Heinemeyer {\it et al.}, 
  JHEP {\bf 0608} (2006) 052
  [arXiv:hep-ph/0604147];
  S.~Heinemeyer, W.~Hollik, A.~M.~Weber and G.~Weiglein,
  JHEP {\bf 0804} (2008) 039
  [arXiv:0710.2972 [hep-ph]].

\bibitem{MicroMegas}
  G.~Belanger, F.~Boudjema, A.~Pukhov and A.~Semenov,
  Comput.\ Phys.\ Commun.\  {\bf 176} (2007) 367
  [arXiv:hep-ph/0607059];
  Comput.\ Phys.\ Commun.\  {\bf 149} (2002) 103
  [arXiv:hep-ph/0112278];
  Comput.\ Phys.\ Commun.\  {\bf 174} (2006) 577
  [arXiv:hep-ph/0405253].

\bibitem{SSARD}  Information about this code is available from K.~A.~Olive: it contains important contributions 
from T.~Falk, A.~Ferstl, G.~Ganis, A.~Mustafayev, J.~McDonald, K.~A.~Olive, P.~Sandick, Y.~Santoso and M.~Srednicki. 

\bibitem{SLHA}
P.~Skands {\it et al.},
  JHEP {\bf 0407} (2004) 036
  [arXiv:hep-ph/0311123];
  B.~Allanach {\it et al.},
  Comput.\ Phys.\ Commun.\  {\bf 180} (2009) 8
  [arXiv:0801.0045 [hep-ph]].


\bibitem{CDFbmm}
T.~Aaltonen {\it et al.}  [CDF Collaboration],
  arXiv:1107.2304 [hep-ex].

\bibitem{JM}
J.~Marrouche, private communication. For a description of {\tt DELPHES}, 
written by S.~Ovyn and X.~Rouby, see
{\tt http://www.fynu.ucl.ac.be/users/s.ovyn/Delphes/index.html}.

\bibitem{XE100}
E.~Aprile {\it et al.}  [XENON100 Collaboration],
  arXiv:1104.2549 [astro-ph.CO].


\end{thebibliography}
\end{document}